# PLATO's signal and noise budget

Anko Börner[1], Carsten Paproth[1], Juan Cabrera[2], Martin Pertenais[1], Heike Rauer[2, 11, 12],
J. Miguel Mas-Hesse[3], Isabella Pagano[4], Jose Lorenzo Alvarez[5], Anders Erikson[2],
Denis Grießbach[1], Yves Levillain[5], Demetrio Magrin[6], Valery Mogulsky[7], Sami-Matias Niemi[5],
Thibaut Prod'homme[5], Sara Regibo[8], Joris De Ridder[8], Steve Rockstein[1], Reza Samadi[9],
Dimitri Serrano-Velarde[7], Alan Smith[10], Peter Verhoeve[5], Dave Walton[10]

**Abstract**

ESA's PLATO mission aims the detection and characterization of terrestrial planets around solar-type stars as well as the study of host star properties. The noise-to-signal ratio (NSR) is the main performance parameter of the PLATO instrument, which consists of 24 Normal Cameras and 2 Fast Cameras. In order to justify, verify and breakdown NSR-relevant requirements the software simulator PINE was developed. PINE models the signal pathway from a target star to the digital output of a camera based on physical models and considers the major noise contributors. In this paper, the simulator's coarse mode is introduced which allows fast performance analyses on instrument level. The added value of PINE is illustrated by exemplary applications.

## 1. Introduction

The PLATO mission (Rauer 2014, Rauer 2016) is the third medium-class mission in ESA's Cosmic Vision programme. Its ambitious goals are the detection and characterization of terrestrial planets around solar-type stars as well as the study of host star properties. The focus of PLATO is on long orbital period terrestrial planets which are also in the habitable zone of solar-type stars. PLATO aims at discovering Earth analog planets under potentially favorable conditions for the development of life.

PLATO will achieve its goals by detecting the low amplitude dips in stellar brightness produced by planets transiting in front of the disk of their parent stars. PLATO has been designed and optimized to continuously monitor stellar fields during several years. The current baseline is a science operations phase of four years which could be distributed in two fields observed for two years each. However, other strategies are possible. It is recalled that PLATO is designed for a 6-year observing duration, the spacecraft will have consumables for up to eight years of operations. To take advantage of these options, mission extensions can be applied for after launch of the mission.

PLATO's science requirement document (PLATO SciRD 2021) defines the top-level goals of the mission. In order to achieve the science goals, stellar samples have been defined for the PLATO observations. Sample 1 (P1) is the backbone of the PLATO mission and must be considered as the highest priority objective. It consists of dwarfs and subgiants with spectral types from F5 to K7 and apparent magnitudes in the V band brighter than $m_V = 11$, observed with a maximum random noise (including photon noise) of $50\ ppm$ in one hour. This noise-to-signal ratio (NSR) is one of PLATO's key performance parameters.

PLATO will employ an array of 24 Normal Cameras and two Fast Cameras (hereafter, 'camera' means the telescope optics and the focal plane assembly, including all the ancillary devices like baffling, electronics, etc.). Each camera is a dioptric telescope with $12\ cm$ entrance pupil feeding four CCDs in their focal plane, each one with $4510 \times 4510\ 18\mu m$ pixels.

The 24 Normal Cameras will provide image data for the primary scientific goals. Their CCDs are Te2v CCD 270 full frame type. Each Normal Camera has a $1037\ deg^2$ effective field-of-view (FoV). The Normal Cameras are optimized to observe stars fainter than $m_V = 8$ with a cadence of $25s$. The 24 Normal Cameras are arranged in four groups of six cameras. All six cameras of each group have exactly the same FoV and pointing. The lines-of-sight of the four groups are offset by a 9.2° angle from the main axis of the payload platform. This particular configuration allows surveying a total field of about $2232\ deg^2$ per pointing, with various parts of the field monitored by 24, 18, 12 or 6 cameras (see Figure 1). This strategy optimizes both the number of targets observed at a given noise level and their brightness. It is assumed that the satellite nominal science orbit will be around the Sun-Earth Lagrange Point 2 (L2). The satellite will be rotated around the mean line of sight by 90° every three months to protect the payload from direct solar radiance during a continuous survey of exactly the same region of the sky.

The CCDs for the Fast Cameras are Te2v CCD 270 frame transfer type meaning that only one half of the CCD acts as photon collector. The other half is termed 'store area'. It is covered and allows a frame transfer operation for rapid charge transfer. This results in a $619\ deg^2$ FoV. The two Fast Cameras will be equipped with a red and a blue filter, respectively, to image stars bi-spectral in the magnitude range from 4 to 8. These cameras work with a cadence of $2.5s$. The images will be used to answer dedicated scientific questions (e.g. Grenfell 2020) and to provide input for high-precision attitude control of the spacecraft.

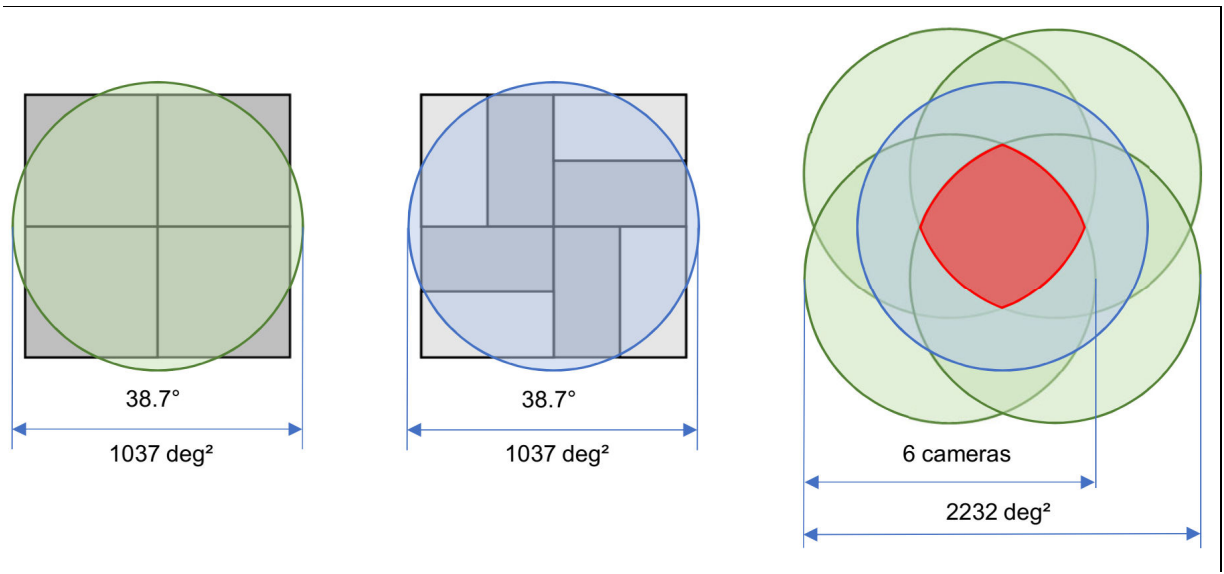

**Figure 1 Normal Camera's field of view (FoV) and focal plane array (FPA) with four full-frame CCDs (left), Fast Cameras FoV and FPA considering frame transfer architecture of the CCDs (center), PLATO instrument FoV depicting the four groups of Normal Cameras in green, the Fast Camera FoV in blue and the jointly covered FoV in red (right). Drawing not to scale.**

The main performance driver for the design of PLATO is to achieve a certain NSR value for a large number of stars. The software simulator PINE (**P**LATO **I**nstrument **N**oise **E**stimator) was developed to predict NSR values of stars in the field of view of PLATO. PINE is used to derive, break down and justify instrument requirements, to perform sensitivity studies, optimizations and trade-offs, to support the generation of PLATO input catalogues (PIC, Montalto 2021) which will be the basic input for PLATO's field selection (Nascimbeni 2022) and to support verification and validation activities during instrument development, characterization/ calibration and in-orbit commissioning. Tools like PINE are an important asset when designing space missions (see also ARIEL radiometric model, Mugnai 2020).

PINE is mainly needed to give a first approximation of the overall PLATO instrument performance, whereas for more detailed simulations, other tools like PLATOSim (Jannsen 2024) are needed. PINE just estimates the overall noise performance with simple models for all the included effects, in contrast to PLATOSim which simulates time-series of CCD pixel values and thus can incorporate many of the same effects in more detail. PINE can give quick answers to overall PLATO performance questions which PLATOSim cannot do without investing a significant amount of resources, but PLATOSim can simulate the detailed behavior of the light curves for single stars which PINE cannot do.

A major challenge is to ensure that the different PLATO simulators produce consistent results as the different simulators use different methods, include different effects, and calculate different outputs. It is an ongoing process to check outputs of different simulators for consistency as the PLATO simulators still evolve, bugs are fixed, and new effects are included. Although it is not part of this paper to compare different simulators, it can be noted that the signal part is identical compared to PLATOSim, while the noise part of PLATOSim has been greatly simplified. However, in the development of the PLATO project, PINE has been the backbone tool against which other simulators (e.g. PLATOSim, but also PSLS, see Samadi 2019) have been tested and validated, in a combined effort including all partners.

## 2. PINE

PINE is a software simulator written in IDL (IDL 2020). It considers all relevant instrument parameters described in PLATO's Technical Requirements Document (PLATO TRD 2022) and characterizes the signal's pathway from a source (here photons from a star) to a digital measurement value through all camera components, e.g. filter, lenses, CCD, electronics. PINE considers all important environmental conditions, e.g. temperature, stray light, radiation and spacecraft jitter. This paper focusses on performance investigations of Normal Cameras in two standard use cases – beginning of life (BOL) and end of life (EOL) with required parameters (the standard instrument parameter sets defined for these use cases). The software can also be applied for different parameter sets, for changed configurations

(e.g. considering the loss of cameras), for Fast Camera simulations (see chapter 5) or for typical parameters. Assuming a normal distribution, "typical" parameters correspond to mean values, whereas "required" parameters are minimum/ maximum values. The main input to PINE is a star magnitude and the main output of PINE is an NSR value for this star, but PINE uses many parameters for its models which needs to be provided as input as well. Simple key value ASCII files are used for the different parameter sets. The parameters PINE uses for its models are described in the next chapter in more detail.

PINE runs in two modes – a **coarse mode** and a **medium mode**. In the coarse mode, simulations focus on a simplified instrument model. Subject of investigation is the joint field of view which is covered by all available Normal Cameras and which is marked in red in Figure 1. It is assumed that all cameras are identical. No spatial variation depending on pixel position is considered for any parameter in the coarse mode, so each pixel at each position in a CCD of each camera will get virtually the same signal (assuming a constant source) and will suffer from the same noise. In this mode, spatial average values are assumed for all parameters varying over a real physical field of view, e.g. transmissivity and charge transfer in-efficiency (CTI). The coarse mode can be considered as a one-dimensional model. Relying on average values is acceptable for dedicated applications of the simulator, since up to 24 Normal Cameras are observing a single star and differences from camera to camera regarding signal and noise are averaged. This mode is used for fast investigations at top mission level. As an example, an output of PINE in the coarse mode is an average NSR of a star with magnitude 11. The coarse mode contains the basic models which are used and extended in the medium mode, but applies these models not to each and every star, separately. This paper introduces PINE's coarse mode in detail.

PINE's medium mode is intended to work on 'object' level, whereby stars or calibration targets are defined as objects. It allows to calculate signal, noise and NSR values for dedicated objects in a certain position of the FoV, e.g. for all stars in a PIC or for a single spot in a thermal-vacuum (TV) chamber, for each camera in any configuration and for each parameter setting. Spatial dependencies of transmissivity, CTI, contamination and quantum efficiency are considered, so the medium mode can be seen as a two-dimensional approach. The medium mode is used to produce new PICs on demand by applying the described noise models to all stars in the PLATO FoV and to verify requirements (e.g. for long-term stability tests). The medium mode will be subject of a later paper.

PINE is considered to be a simulator on system level, where 'system' can be a single camera or the full set of Normal Cameras or even the whole instrument consisting of all available Normal Cameras and the Fast Cameras. Other simulators are developed in parallel for dedicated purposes, i.e. PlatoSim as a pixel-level simulator (Jannsen 2024, comparable to a fictional **fine mode**) or PLATO Solar-like Light-curve Simulator (PSLS) as a data product simulator (Samadi 2019).

Wavelength dependencies are considered in both PINE modes to account for effects in the transmissivity and quantum efficiency. All calculations in all PINE modes can be performed at different time scales, e.g. for $25s$, $600s$, $1h$ and $14h$. The time scales are referring to different scientific goals/ requirements. The input parameters used in PINE vary depending on the use case. In this paper, it is focused on required values of parameters (PLATO TRD 2022) at beginning of life (BOL) and end of life (EOL). Further investigations can be executed on demand, e.g. for 'typical' or 'worst case' parameters, which can be important since the large number of cameras implies a statistical distribution for each parameter or parameters and configurations can change during operation. Just for completeness, PINE version 7.4 with parameter set 1.9 was used for this paper. Both are available from the corresponding author on demand.

## 3. Model description

The camera/ instrument model being the base for coarse and medium mode simulations can be understood as a box containing a number of more or less complex physical or empirical models describing the signal conversion from photons to photoelectrons to digits. Depending on a set of input parameters of the environment, the spacecraft, the optics, the CCDs, the electronics and the operation, this box generates output quantities, such as signal digits and noise digits. The models to derive PLATO's signal and noise budget, that are the building blocks of PINE, are introduced hereafter. A set of real PLATO values is selected to illustrate the approach. These values are required values for Normal Cameras at EOL (PLATO TRD 2022) and can be assumed as a worst-case scenario.

### 3.1. Signal budget

In the signal section of this paper, the pathway of signal photons coming from an observed star is followed through its way of conversion to photoelectrons, voltages and digits based on (simplified) physical models.

We assume that stars emit light as black bodies at a given effective temperature. The flux is calibrated to $3.6182 \cdot 10^{-12}\ W\ cm^{-2}\ \mu m^{-1}$ in V band at $550\ nm$ (Casagrande & VandenBerg 2014). The number of photons is calculated as the radiated power per unit area and unit time, integrating the Planck radiation formula along the wavelength interval of interest, over the energy of a photon of the wavelength of interest (as described below). We have shown in our tests that using spectral energy distributions from spectral libraries (e.g. Coelho 2005) the impact is comparable to the uncertainties that we currently have in the design (e.g. quantum efficiency distribution across the 104 CCD flight models, transmission of the optics, etc.).

### 3.1.1 Camera efficiency

Starting point for building a signal budget is a photon flux density $D_{ph}$ being equivalent to an at-sensor radiance, here assumed for a reference star with magnitude $m_V = 11$. Entering the camera, light passes a cut-off filter and several lenses and windows, which are characterized by on-axis transmissivities $T_{fil}$ and $T_{op}^*$ (including glass, anti-reflective coating, and polarization effects). Both are wavelength depend. In PINE's coarse mode, a single additional wavelength-independent efficiency factor $E_{add}$ of 0.855 is assumed considering contributions due to vignetting (0.91, PLATO TRD 2022, the same source for all other parameters in this section), particulate contamination (0.972), molecular contamination (0.957) and an angle-dependent quantum efficiency (1.01). The latter parameter is bigger than 1.0 since in PLATO most of the light has a large incident angle, whereas the quantum efficiency is determined normally with radiation hitting the CCD surface perpendicularly. All these efficiency factors are averaged values across the cameras FoV. This results in a (wavelength dependent) optical transmissivity $T_{op}$. After that, the wavelength-dependent quantum efficiency of the CCD quantum efficiency ($QE$) has to be considered. All this defines the camera efficiency $E_{cam}$ which allows to calculate the number of generated photoelectrons based on the number of photons received from a target star. Table 3-1Table 3-1 and Figure 2Figure 2 show the parameters of the camera efficiency for BOL and EOL (required values), where $\lambda$ is the wavelength, $D_{ph}^*$ is the photon flux density integrated over the spectral bin size (50 $nm$ in Table 3-1Table 3-1), $QE$ is the quantum efficiency of the CCD, $T_{fil}$ is the transmissivity of the cut-off filters, $T_{op}^*$ is the on-axis transmissivity of the optics, $T_{op}$ is the final optical transmissivity including the additional efficiency factor $E_{add}$. $E_{cam}$ is the wavelength-dependent camera efficiency quantifying the conversion process from photons to photoelectrons.

$$E_{cam}(\lambda) = T_{fil}(\lambda) \cdot T_{op}^*(\lambda) \cdot E_{add} \cdot QE(\lambda) = T_{fil}(\lambda) \cdot T_{op}(\lambda) \cdot QE(\lambda) \qquad \text{Eq. 3-1}$$

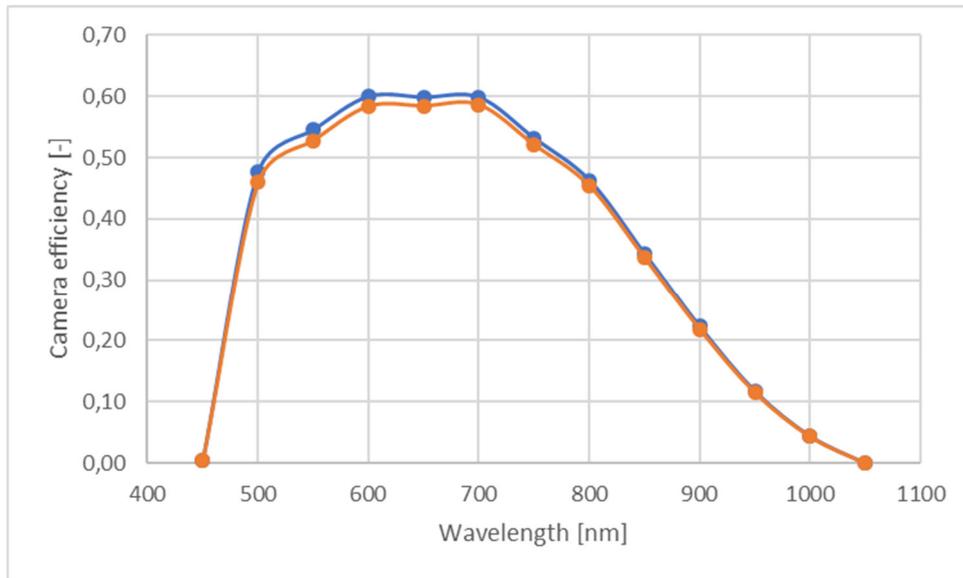

**Figure 2 Camera efficiency requirements for Normal Cameras at BOL (blue) and EOL (orange), to achieve the PLATO science goals.**

|  |  |  | BOL |  |  |  | EOL |  |  |  |
|---|---|---|---|---|---|---|---|---|---|---|
| $\lambda$ | $D_{ph}$* | QE | $T_{fil}$ | $T_{op}^*$ | $T_{op}$ | $E_{cam}$ | $T_{fil}$ | $T_{op}^*$ | $T_{op}$ | $E_{cam}$ |
| [nm] | [ph cm$^{-2}$ s$^{-1}$] | [-] | [-] | [-] | [-] | [-] | [-] | [-] | [-] | [-] |
| 450 | 16,73 | 0,58 | 0,01 | 0,73 | 0,63 | 0,00 | 0,01 | 0,70 | 0,60 | 0,00 |
| 500 | 18,78 | 0,74 | 0,99 | 0,77 | 0,65 | 0,48 | 0,99 | 0,74 | 0,63 | 0,46 |
| 550 | 19,93 | 0,81 | 0,99 | 0,80 | 0,68 | 0,55 | 0,99 | 0,77 | 0,66 | 0,53 |
| 600 | 20,36 | 0,88 | 0,99 | 0,80 | 0,69 | 0,60 | 0,99 | 0,78 | 0,67 | 0,58 |
| 650 | 20,25 | 0,87 | 1,00 | 0,81 | 0,69 | 0,60 | 1,00 | 0,79 | 0,68 | 0,58 |
| 700 | 19,75 | 0,86 | 1,00 | 0,82 | 0,70 | 0,60 | 1,00 | 0,80 | 0,69 | 0,59 |
| 750 | 19,00 | 0,76 | 1,00 | 0,82 | 0,70 | 0,53 | 1,00 | 0,81 | 0,69 | 0,52 |
| 800 | 18,10 | 0,66 | 1,00 | 0,83 | 0,71 | 0,46 | 1,00 | 0,81 | 0,69 | 0,45 |
| 850 | 17,11 | 0,48 | 1,00 | 0,83 | 0,71 | 0,34 | 1,00 | 0,81 | 0,70 | 0,34 |
| 900 | 16,10 | 0,31 | 1,00 | 0,83 | 0,71 | 0,22 | 1,00 | 0,82 | 0,70 | 0,22 |
| 950 | 15,09 | 0,17 | 1,00 | 0,83 | 0,71 | 0,12 | 1,00 | 0,82 | 0,70 | 0,12 |
| 1000 | 14,12 | 0,06 | 1,00 | 0,83 | 0,71 | 0,04 | 1,00 | 0,82 | 0,70 | 0,04 |
| 1050 | 13,18 | 0,00 | 0,96 | 0,84 | 0,72 | 0,00 | 0,96 | 0,82 | 0,70 | 0,00 |

**Table 3-1 Efficiency parameters requirements for Normal Cameras at BOL and EOL.**

### 3.1.2 Photon-photoelectron conversion

Based on the number of photons entering the camera from an object, the total number of photoelectrons generated in the CCD $n_{pe\_t\_total}$ can be calculated as

$$n_{pe\_t\_total} = A_{ap} \cdot t_{exp} \cdot E_{add} \cdot \int_{\lambda 1}^{\lambda 2} D_{ph}(\lambda) \cdot T_{op}^*(\lambda) \cdot T_{fil}(\lambda) \cdot QE(\lambda) \, d\lambda \qquad \text{Eq. 3-2}$$

with

- $D_{ph}(\lambda)$ — spectral photon flux density of a target star, $[\frac{ph}{cm^2 \cdot s \cdot nm}]$
- $\lambda$ — wavelength, $[nm]$
- $\lambda 1$ — lower limit set for the simulations, $450nm$
- $\lambda 2$ — upper limit set for the simulations, $1050nm$
- $A_{ap}$ — effective entrance aperture of the optics, $\pi/4 \cdot 12 \, cm^2 = 113.1 cm^2$
- $t_{exp}$ — exposure time (static integration time), $21s$ for Normal Cameras, cadence of $25s$
- $E_{add}$ — additional wavelength-independent efficiency factor, $0.855$
- $T_{op}^*(\lambda)$ — on-axis transmissivity of the optics (glass, coating, polarization), see Table 3-1Table 3-1
- $T_{fil}(\lambda)$ — transmissivity of the filters, see Table 3-1Table 3-1
- $QE(\lambda)$ — quantum efficiency of the CCD, see Table 3-1Table 3-1

Assuming a solar-type star with $m_V = 11$ and the given required PLATO parameters, the $508000$ photons entering the camera over the entire wavelength range within one exposure are converted into about $192000$ photoelectrons. Again, it shall be mentioned, that this value is an average number for a single star in a single image of a single camera.

### 3.1.3 Imagettes and masks

Because of telemetry limitations the PLATO mission cannot download full-frame images to ground continuously. In order to overcome this limitation, the payload Data Processing Unit (DPU) will process onboard the CCD windows ("imagettes") around each target star. Each imagette has a typical size of $6 \times 6$ pixels, which could be enlarged for saturated stars. For all P1 sample stars, the imagettes will be downloaded to ground without further processing. On ground, a (weighted) photometric mask can be applied on imagettes to optimize the noise-to-signal ratio (NSR) of the investigated star. For fainter targets ($V < 13$), the total flux within an imagette will be calculated on-board, using binary masks at a nominal cadence of $600s$ (24 acquisitions of $25s$), but with the possibility of acquiring flux values at $50s$ for a reduced number of targets to improve NSR.

### 3.1.4 Point spread function and phase function

The incoming point-source signal is spread over a number of pixels due to system point spread function (PSF, here considering optics and CCD charge diffusion) and a phase function. In PINE, a Gaussian shaped PSF kernel is assumed with a size of $\sigma = 0.64\,pix$ (approximated from ZEMAX simulations of the PLATO optics) which corresponds to an enclosed energy of 32% in a perfectly centered pixel. The phase function, see Figure 3, considers that some of the stars will be projected into the center of a pixel, whereas others will be projected to the pixel's edge and integrating the PSF over a pixel will thus create different signal levels in a pixel depending on the phase function.

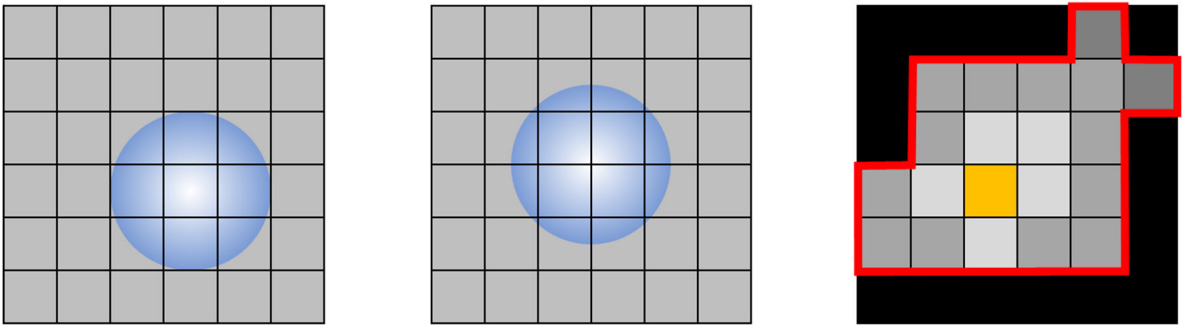

**Figure 3 Concept of phase function (where the left image shows a star centered and the center image shows a star laying at pixels edge) and an imagette (here 6 by 6 pixels) with a masked area (outlined in red) and the brightest pixel in orange (right). Note that the blacked-out area will not be used to calculate the stellar flux.**

In average, the phase function brings the 32% enclosed energy down to 27% for the brightest pixel in the mask which needs to be considered applying a statistical approach such as for PINE's coarse mode. This fraction of the enclosed energy is called here the signal spread factor $f_{sp}$. The rest of the energy (73%) is spread over neighboring pixels. Signal spread has an impact to final NSR due to additional noise and signal-dependent CTI effects. Considering $f_{sp}$, the signal of the brightest pixel in an imagette caused by a target star can be determined as:

$$n_{pe\_t} = n_{pe\_t\_total} \cdot f_{sp} \qquad \text{Eq. 3-3}$$

While $n_{pe\_t}$ is the number of photoelectrons in the brightest pixel, $n_{pe\_t\_total}$ is the number of photoelectrons in the integrated PSF. In PINE, $n_{pe\_t}$ is relevant for considerations on a single pixel (camera level) whereas $n_{pe\_t\_total}$ is relevant for considerations on the masked area on instrument level.

### 3.1.5 Background, stray light, contaminating stars

Beyond the wanted signal from the object, additional photons are collected from background, stray light from spacecraft and instrument (total: $124\,\frac{e^-}{pix\,s}$, background: $100\,\frac{e^-}{pix\,s}$, spacecraft: $4\,\frac{e^-}{pix\,s}$, instrument: $20\,\frac{e^-}{pix\,s}$), and contaminating stars. The contribution of the latter was estimated by evaluating the PPMXL star catalogue (Roeser 2010) and found to be 1% of the target star signal (median value). These additional photoelectrons are summed up and contribute to the number of photoelectrons in the brightest pixel $n_{pe}$:

$$n_{pe} = n_{pe\_t} + n_{pe\_b} + n_{pe\_s} + n_{pe\_c} \qquad \text{Eq. 3-4}$$

with

    $n_{pe\_b}$ - photoelectrons from background (estimated direct imaging of Zodiacal light), $60\,\frac{e^-}{pix\,s}$

    $n_{pe\_s}$ - photoelectrons from stray light from spacecraft and cameras, $64\,\frac{e^-}{pix\,s}$

    $n_{pe\_c}$ - photoelectrons from contaminating stars, 1% of the target star signal

### 3.1.6 Charge transfer efficiency

CCDs are light sensitive shift registers. Charges generated in a pixel are transferred through all pixels in the same column between that pixel and a readout register in a parallel transfer process. In that register the charges are transferred serially to a readout gate. Both charge transfer processes, through all pixels and from the register, have different pixel-to-pixel transfer efficiencies/ in-efficiencies. This leads to an additional signal spread in transfer direction and to a reduction of the signal level in the directly illuminated pixels. PLATO CCD270 detectors have a size of $4510 \times 4510$ pixels. There are two readout registers per CCD, each managing one half CCD. A full transfer includes then 4510 parallel transfers and 2255 serial transfers. The average transfer efficiency for a CCD $E_{CTI}$ can be estimated as:

$$E_{CTI} = \frac{1}{IJ} \sum_{j=1}^{J} \sum_{i=1}^{I} (1 - CTI_{par})^j (1 - CTI_{ser})^i \qquad \text{Eq. 3-5}$$

with

$CTI_{par}$ - charge transfer in-efficiency for parallel readout, $[-]$
$CTI_{ser}$ - charge transfer in-efficiency for serial readout, $[-]$
$I$ - number of columns, $[-]$
$J$ - number of rows, $[-]$

In PINE's coarse mode, the CTI caused efficiency is set to an average value considering the size of the CCD. Based on ESA measurements (Prod'homme 2016) an empirical model was derived for different radiation levels.

$$CTI_{ser} = 0.1 \cdot CTI_{par} \qquad \text{Eq. 3-6}$$

$$CTI_{par} = n_{pe}{}^m \cdot 10^n \qquad \text{Eq. 3-7}$$

with

$n_{pe}$ - number of photoelectrons per pixel for a target star, depending on camera efficiency and signal spread, $[-]$
$m$ - constant parameter, $-0.6541$
$n$ - radiation dependent parameter

with
$$n = f(R) = a \cdot R^2 + b \cdot R + c \qquad \text{Eq. 3-8}$$

$R$ - radiation level, total non-ionizing dose, $10MeV$, $1.73e9$ $[p^+/cm^2]$
$a$ - model parameter, $-3.2110 \cdot 10^{-20}$
$b$ - model parameter, $3.6124 \cdot 10^{-10}$
$c$ - model parameter, $-1.9103$

The applied radiation level is derived from an analysis (Verhoeve 2016) which showed at CCD level a maximum total non-ionizing dose (TNID) of $5e9$ $[p^+/cm^2]$ and a mean TNID of $2.5e9$ $[p^+/cm^2]$ for 6.5 years. For evaluating PLATO's performance, EOL was scaled down to 4.5 years. The CTI model allows to estimate the number of photoelectrons at the readout gate of the CCD after parallel and serial charge transfers.

$$n_{pe\_CTI} = n_{pe\_t} \cdot E_{CTI} \qquad \text{Eq. 3-9}$$

The charges lost due to the inefficiency of the transfer reduce the signal level in the PINE model. It is subject to random distribution, but are not considered as an additional noise source for stars.

### 3.1.7 Charge-to-voltage conversion

The photoelectrons transferred through parallel and serial shift registers are converted into a voltage at the CCD readout register.

$$V_{CCD} = n_{pe\_CTI} \cdot C(T,V) \qquad \text{Eq. 3-10}$$

with

$n_{pe\_CTI}$ - number of photoelectrons generated in the detector element and transferred to the readout register, $[-]$
$C(T,V)$ - charge-voltage conversion factor, depending on temperature and bias voltages, $1.8\mu V/e^-$, here assumed to be constant

Temperature and bias voltage dependence of the charge-voltage conversion has to be further analyzed during the camera test campaign and the model has to be updated accordingly.

In PLATO's front-end electronics (FEE), incoming CCD signals are sampled, amplified, and finally digitized ($16 bit$). FEE offset and FEE gain convert the incoming signal into the output signal $DN_{FEE}$.

$$DN_{FEE} = V_{CCD} \cdot g_{FEE} + o_{FEE}$$

Eq. 3-11

with

$g_{FEE}$ - FEE gain, $0.0222 \, ADU/\mu V$
$o_{FEE}$ - FEE offset, $500 \, ADU$

The gain is set to match the dynamic range, the FEE offset will have to be adapted during assembly and electrical commissioning of the spacecraft in orbit. The parameters $g_{FEE}$ and $o_{FEE}$ quoted above are representative of the current PLATO design. However, the final in-flight values may be different for each camera (and can be adapted later on demand).

The entire process of signal transfer is illustrated for an example in Table 3-2Table 3-2, where a $m_V = 11$ target star is projected to a single Normal Camera (EOL, required input parameters). Numbers in the table hereafter are rounded.

| Key parameter | Variable | Value EOL |
|---|---|---|
| Number of photons coming from one target star ($mv = 11$) entering the entrance pupil of PLATO's optics within one exposure in one camera between $450$ and $1050 nm$ | $n_{ph\_t}$ | 508000ph |
| Mean efficiency between $500$ and $950 nm$ including, optics (glass, vignetting, polarization, anti-reflection (AR) coating), filter, molecular and particulate contamination, QE, angle dependent QE | $E_{misc}$ | 0.38 |
| Total number of generated photoelectrons from the target star | $n_{pe\_t\_total}$ | 192000$e^-$ |
| Signal spread factor | $f_{sp}$ | 0.27 |
| Number of photoelectrons in the brightest pixel within a mask generated from the target star | $n_{pe\_t}$ | 51200$e^-$ |
| Number of photoelectrons in the brightest pixel considering additional background signal, photons from contaminating stars and stray light | $n_{pe}$ | 54400$e^-$ |
| Parallel CTI | $CTI_{par}$ | $3.3 \cdot 10^{-5}$ |
| Efficiency due to CTI | $E_{cti}$ | 0.93 |
| Number of photoelectrons (from target star, background, stray light, contaminating stars) after considering CTI of the CCD | $n_{pe\_CTI}$ | 50300$e^-$ |
| Number of photoelectrons in the brightest pixel coming from the target star only considering CTI of the CCD. This value is taken as reference for NSR calculations on camera level | $n_{ref\_p}$ | 47400$e^-$ |
| Number of photoelectrons collected in the entire mask | $n_{ref\_m}$ | 177100$e^-$ |
| CCD output voltage | $V_{CCD}$ | $90.5 mV$ |
| FEE output signal | $DN_{FEE}$ | 2510ADU |

**Table 3-2 Example for signal conversion in a single PLATO Normal Camera**

## 3.2. Noise budget

PINE's modular concept allows to integrate various noise components or noise contributors. A noise-to-signal ratio (NSR) is calculated for each contributor, for different time scales (e.g. $25 s$ or $1 h$) and on two levels – on camera level and on instrument level. NSR on camera level $NSR^C$ describes quantities for a single (brightest) pixel of a single camera over different time scales which is important for verifications tasks in thermal-vacuum (TV) chambers, for example. NSR on instrument level $NSR^I$ considers the entire signal in the mask (imagette) and the related noise. It applies to all available ($24$ at BOL, $22$ at EOL) Normal Cameras ("instrument") combined, and includes additional instrument related noise effects, e.g. electro-magnetic compatibility (EMC) and spacecraft pointing jitter noise. NSR on instrument level is calculated for $1 h$ only, because $1 h$ is used for the PLATO NSR requirements.

For PLATO, noise is generally divided in three main categories – photon noise from the target star (which shall always be dominant for a reference star), noise from random sources (called random noise) and systematic noise.

Systematic effects mostly depend on temperature differences (at optics, CCD, FEE) and CCD bias voltages. On larger time scales, differential kinematic aberration will play an important role and these drift effects need to be considered (Samadi 2019). This paper focusses on short-term performance issues only, therefore it is assumed that systematics will be corrected based on auxiliary measurements (e.g. CCD temperature measurements) and the residuals are Gaussian distributed. But systematic noise contributions which depend on the same physical effect (e.g. optics' temperature) do not behave independently, so the overall noise accumulation needs to consider this accordingly (see chapter 4). In total, PINE considers more than 20 noise sources.

The noise section focuses on the most relevant contributors only (3.2.1 - 3.2.8). It shall be noted, that different noise contributors can be dominant or non-relevant at different times (e.g. BOL, EOL), at different time scales (e.g. $25s$, $14h$) and/ or at the two different levels (camera or instrument). Below, the quantities being important for the $1h$ observation period (PLATO SciRD 2021) are examined in more detail.

### 3.2.1 Photon noise target star

Photon noise is a signal-inherent noise source. On instrument level it can be averaged over time and over the number of cameras. For the target star it is estimated as

$$NSR_{ph}^C = \frac{1}{\sqrt{n_{ref\_p}}} \cdot 1e6\ ppm \qquad \text{Eq. 3-12}$$

$$NSR_{ph}^I = \frac{1}{\sqrt{n_{ref\_m}}} \cdot \frac{1e6\ ppm}{\sqrt{n_{cam}} \cdot \sqrt{n_{img}}} = \frac{\sqrt{f_{sp}}}{\sqrt{n_{ref\_p} \cdot n_{cam} \cdot n_{img}}} \cdot 1e6\ ppm \qquad \text{Eq. 3-13}$$

with

| | |
|---|---|
| $n_{ref\_p}$ | - number of photoelectrons from a reference star in the brightest pixel, [-] |
| $n_{ref\_m}$ | - number of photoelectrons from a reference star in the entire mask, [-] |
| $n_{cam}$ | - number of cameras, 22 |
| $n_{img}$ | - the number of images taken within $1h$, 144 |

The 22 Normal Cameras are a worst-case estimation at EOL based on a reliability analysis and the assumption that two of the 24 Normal Cameras are not operating any longer. 144 images result from a cadence of $25s$.

### 3.2.2 CCD readout noise

CCD readout noise is caused by the output amplifier on the CCD. Noise-to-signal ratio on camera level and on instrument level can be calculated as

$$NSR_{CCD\_ro}^C = \frac{err_{CCD\_ro}}{n_{ref\_p}} \cdot 1e6\ ppm \qquad \text{Eq. 3-14}$$

$$NSR_{CCD\_ro}^I = \frac{err_{CCD\_ro}}{n_{ref\_m}} \cdot \frac{1e6\ ppm \cdot \sqrt{n_{mask}}}{\sqrt{n_{cam}} \cdot \sqrt{n_{img}}} \qquad \text{Eq. 3-15}$$

$$= \frac{err_{CCD\_ro} \cdot f_{sp} \cdot \sqrt{n_{mask}}}{n_{ref\_p} \cdot \sqrt{n_{cam} \cdot n_{img}}} \cdot 1e6\ ppm$$

with

| | |
|---|---|
| $err_{CCD\_ro}$ | - readout noise of the CCD, $44.3e^-$ |
| $n_{mask}$ | - number of noise effective mask pixels |

Since, on instrument level the entire signal in the mask is considered, also more than one pixel contributes to the noise. An optimal mask size and shape depends strongly on the magnitude of the target star, its position in the focal plane and the presence of contaminating stars (Marchiori 2019). This paper aims to verify top-level requirements on instrument level (performance for stars with $m_V = 11$, NSR $50 ppm$ in $1hr$), so a simplified approach is followed illustrating the general methodology. Using an empirical model, the number of noise effective pixels in the mask $n_{mask}$ was determined to be $9.5\ pix$ for a reference star.

### 3.2.3 FEE readout noise

FEE readout noise is caused by the output amplifier on the front-end electronics. Noise-to-signal ratio on camera level and on instrument level can be calculated as

$$NSR_{FEE\_ro}^{C} = \frac{err_{FEE\_ro}}{n_{ref\_p}} \cdot 1e6\ ppm \qquad \text{Eq. 3-16}$$

$$NSR_{FEE\_ro}^{I} = \frac{err_{FEE\_ro}}{n_{ref\_m}} \cdot \frac{1e6\ ppm \cdot \sqrt{n_{mask}}}{\sqrt{n_{cam}} \cdot \sqrt{n_{img}}} = \frac{err_{FEE\_ro} \cdot f_{sp} \cdot \sqrt{n_{mask}}}{n_{ref\_p} \cdot \sqrt{n_{cam} \cdot n_{img}}} \cdot 1e6\ ppm \qquad \text{Eq. 3-17}$$

with

$err_{FEE\_ro}$ — readout noise of the FEE, $37.0 e^{-}$

### 3.2.4 Spacecraft pointing jitter noise

The spacecraft (S/C) pointing jitter will induce noise in the light curves. This noise is systematic, it is uncorrelated to other noise sources and it strongly depends on the shape and size of the aperture mask. Predicting the impact of S/C jitter is a complicated task and the most accurate approaches involve pixel-level simulators (like PLATOSim, Jannsen 2024) and simulated high-frequency pointing time series. The impact of jitter noise can be mitigated by a jitter correction approach (e.g. Fialho 2007), the PLATO on-ground data reduction pipeline already includes this possibility. However, for PLATO it was decided at the initial stage of the mission to include a S/C jitter noise budget allocation at system level. This allocation is based on the best knowledge at the time of the design of the mission, the pointing requirements on the attitude and orbital control system of the spacecraft, and the performance of the fine guidance system (Grießbach 2020). The noise budget for residuals of jitter noise after correction was set to $9 ppm$ in one hour on instrument level (PLATO TRD 2022) and it is the major contributor to the overall systematic noise budget of $11.3 ppm$. These values can be converted into an amplitude spectral density (ASD) function (0.54 and 0.68 $ppm/\sqrt{\mu Hz}$, respectively, see Figure 4 from PLATO Red Book 2017). At frequencies below $20 \mu Hz$, which correspond to a time duration greater than 14 hours, or time scales shorter than the transit of an Earth-size planet orbiting at 1 AU around a G type star, the requirement is defined by the need to study oscillations of several categories of early-type stars. The noise level requirement in these cases is less stringent with decreasing frequency, as the maximum mode height of the oscillations increases. As a consequence, the mean residual error is allowed to rise gradually from its specification at $20 \mu Hz$ to lower frequencies, reaching a maximum of $50 ppm/\sqrt{\mu Hz}$ at a frequency of $3 \mu Hz$ (or ~4 days) for stars with $mV = 11$.

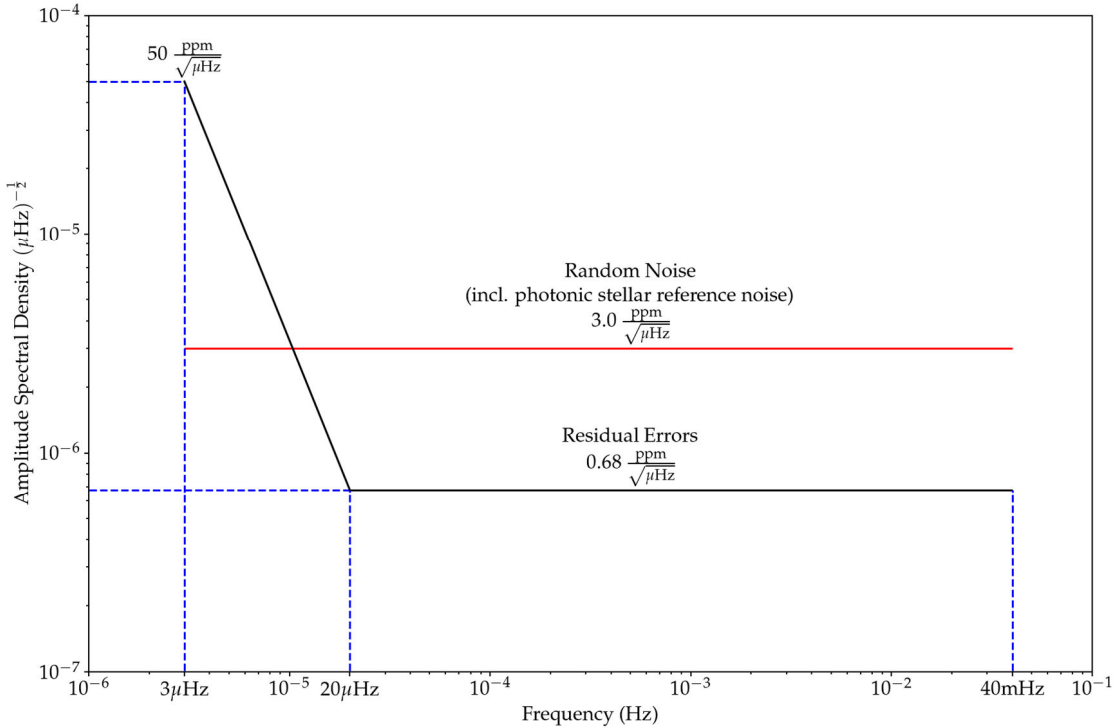

**Figure 4 ASD of required random and jitter noise for PLATO. At the low frequency side, a higher systematic error of up to 50 $ppm/\sqrt{\mu Hz}$ is allowed to account for difficult to correct slow drifts which aren't confused as stellar spectral features.**

NSR on camera level $NSR^{C}{}_{jitter}$ and on NSR on instrument level $NSR^{I}{}_{jitter}$ are then defined as:

$$NSR^C_{jitter} = ASD_{jitter}(f_{cadence}) \cdot \sqrt{f_{cadence}} = 108\text{ppm} \qquad \text{Eq. 3-18}$$

$$NSR^I_{jitter} = ASD_{jitter}(f_{obs}) \cdot \sqrt{f_{obs}} = 9\text{ppm} \qquad \text{Eq. 3-19}$$

with

$f_{cadence}$ - cadence frequency, $40000\mu Hz$ for Normal Camera with cadence $25s$
$f_{obs}$ - observation frequency, $278\mu Hz$ for observation time $1h$

### 3.2.5 Miscellaneous photon noise

Beyond photon noise from the target star, which has a separate budget and which is dominating the overall NSR, there are several light sources contributing to signal and noise. These sources cause so-called miscellaneous photon noise. The contributors are described in 3.1.5. Noise-to-signal ratio on camera level and on instrument level can be calculated as

$$NSR^C_{misc} = \frac{\sqrt{n_{pe\_b} + n_{pe\_c} + n_{pe\_s}}}{n_{ref\_p}} \cdot 1e6 \; ppm \qquad \text{Eq. 3-20}$$

$$NSR^I_{misc} = \frac{\sqrt{n_{pe\_b} + n_{pe\_c} + n_{pe\_s}}}{n_{ref\_m}} \cdot \frac{\sqrt{n_{mask}}}{\sqrt{n_{cam} \cdot n_{img}}} \cdot 1e6 \; ppm \qquad \text{Eq. 3-21}$$

The NSR on instrument level will be improved by averaging over the number of cameras and the number of images. All pixels in the mask have to be considered.

### 3.2.6 CCD smearing photon noise

The readout mechanism of the CCDs being applied for PLATO cameras will lead to a smearing of a signal over a complete CCD column. During a transfer of the signal of a target star in the parallel readout register, the moving package of collected charges will be "enriched" by photoelectrons from other stars illuminating pixels in the same column. The signal itself is an offset, but its photon noise needs to be considered. Noise-to-signal ratio on camera level and on instrument level can be calculated as

$$NSR^C_{smear} = \frac{\sqrt{r_{smear} \cdot t_{trans}}}{n_{ref\_p}} \cdot 1e6 \; ppm \qquad \text{Eq. 3-22}$$

$$NSR^I_{smear} = \frac{\sqrt{r_{smear} \cdot t_{trans}}}{n_{ref\_m}} \cdot \frac{\sqrt{n_{mask}}}{\sqrt{n_{cam} \cdot n_{img}}} \cdot 1e6 \; ppm \qquad \text{Eq. 3-23}$$

with

$r_{smear}$ - median photoelectron rate, $45 \frac{e^-}{s\,pix}$ (estimation based on the catalog PPMXL, Roeser 2010)
$t_{trans}$ - charge transfer time, difference between cadence and exposure time, $4s$

The NSR on instrument level will be improved by averaging over the number of cameras and the number of images. All pixels in the mask have to be considered.

### 3.2.7 PSF breathing noise

Temperature variations of the telescope cause changes of the imaging properties of the telescope (in particular the focus), resulting in a change of the size and the shape of the PSF. A changing PSF leads to variations in the measured flux, at least if single pixels are considered. PSF breathing noise depends on:
- amplitude and frequency of temperature variations of the telescope,
- size and weights of the photometric mask,
- local photo-response non-uniformity (PRNU, low impact),
- dark signal non-uniformity (DSNU, very low impact),
- thermo-elastic stability of the FPA (very low impact).

Temperature is measured and controlled independently for each telescope but the accuracy of temperature control is limited. The effects of temperature changes on the measured flux in the relevant detector pixels can be modelled and this allows a partial correction of the systematic error. Residuals are assigned to a noise contributor. The NSR values caused by PSF breathing noise are budgeted in PLATO's requirement document (PLATO TRD 2022). Accumulating over the number of cameras and several images will improve NSR on instrument level since breathing of different telescopes is not correlated due to independent temperature controls.

$$NSR^C_{psf\_breath} = 20\ ppm\ \text{(PLATO TRD 2022)}$$ **Eq. 3-24**

$$NSR^I_{psf\_breath} = \frac{20 ppm}{\sqrt{n_{cam} \cdot n_{img}}}$$ **Eq. 3-25**

### 3.2.8 FEE offset stability noise

FEE offset (or bias) varies with temperature. It is linked directly to the stability (on camera level) and the measurement accuracy (on instrument level) of the FEE temperature. Noise-to-signal ratio on camera level and on instrument level can be calculated as

$$NSR^C_{FEE\_off} = \frac{err_{FEE\_OS} \cdot G_{ADC} \cdot T_{FEE}}{n_{ref\_p}} \cdot f_{conv} \cdot 1e6\ ppm$$ **Eq. 3-26**

$$NSR^I_{FEE\_off} = \frac{err_{FEE\_OS} \cdot G_{ADC} \cdot T_{meas}}{n_{ref\_m}} \cdot \frac{f_{conv}}{\sqrt{n_{cam} \cdot n_{img}}} \cdot 1e6\ ppm$$ **Eq. 3-27**

with

| | |
|---|---|
| $err_{FEE\_OS}$ | - temperature dependent offset sensitivity, $1\ \frac{ADU}{K}$ |
| $G_{ADC}$ | - ADC gain, $25\ \frac{e^-}{ADU}$ |
| $T_{FEE}$ | - temperature stability for FEE, $\pm 1.0 K$ |
| $f_{conv}$ | - conversion factor from peak value to 1 sigma, 2/3 |
| $T_{meas}$ | - temperature knowledge, $0.01 K$ |

Since temperature control is independent for each camera, averaging the signal over the number of images and the number of cameras will improve NSR. FEE offset stability affects all pixels of the photometric mask the same way, so no additional factor has to be applied.

## 4. Overall NSR

The final step is the accumulation of all single noise contributions. As mentioned in the beginning, three major noise groups are considered - photon noise from target star $NSR_{ph}$, random noise $NSR_{rnd}$ and systematic noise $NSR_{sys}$. The photon noise is expected to be the main contributor and PINE uses only simple models to estimate other noise contributors. Note that in our approximation the jitter noise plays the role of a noise floor which is commonly used to account for low-frequency instabilities and de-correlation residuals (e.g. Mugnai 2020 and references therein). The overall NSR can be determined as

$$NSR_{overall} = \sqrt{NSR_{ph}^2 + NSR_{rnd}^2 + NSR_{sys}^2}$$ **Eq. 4-1**

Each of the noise components is labeled according to its category. $NSR_{rnd}$ can be calculated by the square root of the $K1$ summed squared individual random NSR contributions (marked with "r" in the "noise type" column in Table 5-1Table 5-1).

$$NSR_{rnd} = \sqrt{\sum_{k=1}^{K1} NSR_k^2}$$ **Eq. 4-2**

If a noise source is labelled as 'systematic' then it is either related to one of the camera temperatures or it is not correlated. $NSR_{sys}$ can be expressed by four independent terms:

$$NSR_{sys} = \sqrt{NSR_u^2 + NSR_{T\_TOU}^2 + NSR_{T\_CCD}^2 + NSR_{T\_FEE}^2}$$ **Eq. 4-3**

where $NSR_u$ describes uncorrelated terms (e.g. spacecraft pointing jitter, aberration noise). Since these terms are independent, the $K2$ uncorrelated contributors can by summarized this way:

$$NSR_u = \sqrt{\sum_{k=1}^{K2} NSR_k^2}$$ **Eq. 4-4**

If $K3$ NSR terms are correlated with the optics temperature then they need to be simply summed:

$$NSR_{T\_TOU} = \sum_{k=1}^{K3} NSR_k \qquad \text{Eq. 4-5}$$

The same approach applies for NSR terms depending on the CCD temperature $NSR_{T\_CCD}$ or on the FEE temperature $NSR_{T\_FEE}$.

## 5. Application of PINE

This section illustrates different applications of PINE in its coarse mode.

### 5.1. Top-level requirement verification

The described coarse PINE model was used to estimate PLATO's NSR on instrument level and to validate the system design w.r.t. top-level system requirements. Two use cases are of special importance – PLATO's estimated performance with required parameters at beginning of life (BOL) and at end of life (EOL after 4.5 years). The main differences between performance on instrument level at BOL and EOL are caused by the assumed loss of two cameras (based on a risk assessment) and by the CTI degradation on CCD level due to accumulated total non-ionizing dose (TNID) radiance. By doing this investigation the PLATO consortium proved that the system design and several assumed, measured, or required parameters of units or components allow to fulfill the top-level science goals of PLATO. Table 5-1Table 5-1 shows the results for these two use cases for Normal Camera for reference stars ($m_V = 11$).

The combined random and photon noise is $49.3 ppm$ (highlighted in green) and, therefore, below the required $50 ppm$ value even for worst-case estimations at EOL. At BOL PLATO will deliver data products with $44.3 ppm$ for random noise and photon noise (highlighted in blue). These values confirm that the top-level scientific requirements (PLATO SciRD 2021) can be fulfilled. The overall noise including the residuals of the corrected systematics (photon noise, random noise and systematic noise) will be $45.2 ppm$ and $50.1 ppm$, respectively, on instrument level.

| Noise contributor | Noise type | Corr. type | BOL Camera 1img | BOL Camera 1hr | BOL Camera 14hr | BOL Instr. 1hr | EOL Camera 1img | EOL Camera 1hr | EOL Camera 14hr | EOL Instr. 1hr |
|---|---|---|---|---|---|---|---|---|---|---|
| Photon noise target star | r | u | 4414,5 | 367,9 | 98,3 | 38,8 | 4592,6 | 382,7 | 102,3 | 42,2 |
| Spacecraft pointing jitter noise | s | u | 108,0 | 9,0 | 2,4 | 9,0 | 108,0 | 9,0 | 2,4 | 9,0 |
| PSF breathing noise | s | t_o | 20,0 | 20,0 | 40,0 | 0,3 | 20,0 | 20,0 | 40,0 | 0,4 |
| Miscellaneous photon noise | r | u | 1088,0 | 90,7 | 24,2 | 15,2 | 1170,2 | 97,5 | 26,1 | 17,1 |
| CCD readout noise | r | u | 754,2 | 62,8 | 16,8 | 10,5 | 934,4 | 77,9 | 20,8 | 13,7 |
| CCD smearing photon noise | r | u | 261,5 | 21,8 | 5,8 | 3,7 | 283,0 | 23,6 | 6,3 | 4,1 |
| FEE offset stability noise | s | t_f | 324,8 | 27,1 | 7,2 | 0,1 | 351,5 | 29,3 | 7,8 | 0,1 |
| FEE readout noise | r | u | 623,6 | 52,0 | 13,9 | 8,7 | 780,4 | 65,0 | 17,4 | 11,4 |
| Photon noise | | | 4414,5 | 367,9 | 98,3 | 38,8 | 4592,6 | 382,7 | 102,3 | 42,2 |
| Random noise | | | 1500,3 | 125,0 | 33,8 | 21,4 | 1736,8 | 144,7 | 39,0 | 25,4 |
| Photon and random noise (top level requirement) | | | 4662,5 | 388,5 | 104,0 | 44,3 | 4910,0 | 409,2 | 109,5 | 49,3 |
| Systematic noise | | | 542,2 | 50,0 | 43,6 | 9,0 | 568,2 | 51,9 | 43,8 | 9,0 |
| Overall noise | | | 4693,9 | 391,7 | 112,7 | 45,2 | 4942,8 | 412,5 | 117,9 | 50,1 |

Table 5-1 Estimated NSR values for PLATO's Normal Cameras, at BOL and EOL, required values, $m_V = 11$ on camera level and on instrument level for different time scales. All numbers in [ppm]. Noise types are (r)andom or (s)ystematic, correlation types are (u)ncorrelated or correlated via optics (t_o) or FEE (t_f)

### 5.2. Investigation of different target stars

PLATO will observe about $5.2\%$ of the sky with approximately $200000$ stars. In order to estimate the NSR not only of reference stars but also of brighter and fainter stars, the major noise categories were investigated depending on the magnitude of the stars. It shall be mentioned again, that in PLATO noise is divided in three major categories – photon noise from target star, noise from random sources (called random noise) and systematic noise. The results of this investigation are shown in a log plot in Figure 5. The sum of photon noise and random noise is highlighted due to its relevance for requirement verification.

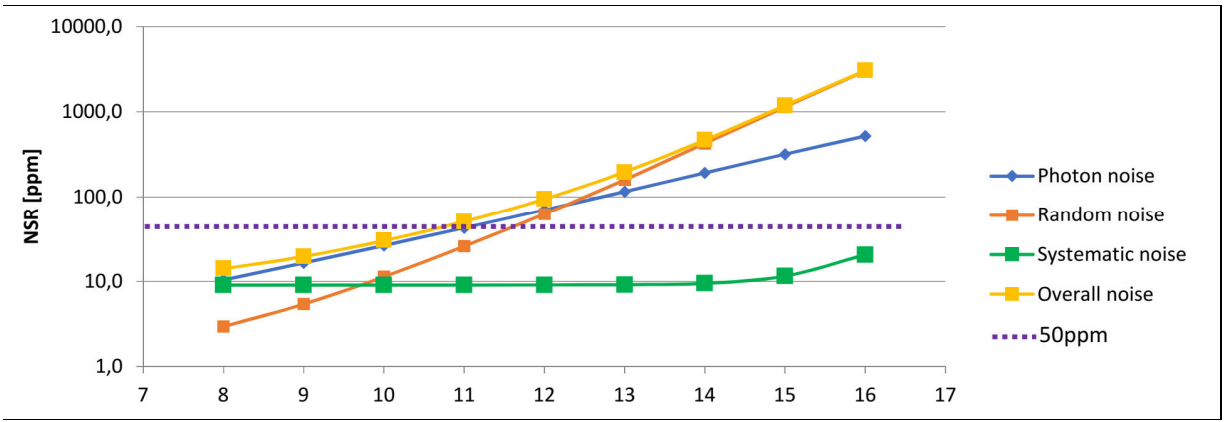

**Figure 5 Estimated NSR values for PLATO's Normal Camera, EOL, $m_V = 8 \ldots 16$, log plot**

| Magnitude | [-] | 8 | 9 | 10 | 11 | 12 | 13 | 14 | 15 | 16 |
|---|---|---|---|---|---|---|---|---|---|---|
| Photon noise | [ppm] | 10,3 | 16,3 | 26,2 | 42,2 | 68,9 | 114,0 | 190,1 | 315,0 | 513,7 |
| Random noise | [ppm] | 2,2 | 4,5 | 10,3 | 25,4 | 66,0 | 178,7 | 495,1 | 1360,6 | 3625,8 |
| **Photon and random noise** | [ppm] | 10,5 | 17,0 | 28,1 | 49,3 | 95,4 | 211,9 | 530,3 | 1396,6 | 3662,0 |
| Systematic noise | [ppm] | 9,0 | 9,0 | 9,0 | 9,0 | 9,0 | 9,1 | 9,4 | 11,3 | 20,1 |
| Overall noise | [ppm] | 13,9 | 19,2 | 29,5 | 50,1 | 95,8 | 212,1 | 530,4 | 1396,6 | 3662,0 |

**Table 5-2 Estimated NSR values for PLATO's Normal Camera, EOL, $m_V = 8 \ldots 16$**

From this analysis, it can be seen that the top-level requirement (the NSR of a star with a magnitude of 11 must not exceed $50 ppm$) can be fulfilled.

### 5.3. Sensitivity analyses

In general, one main purpose of simulators is to provide sensitivity analyses. It is often of interest to answer questions like "What happens to an error metric, if parameter X changes?" or "What is the best trade-off to compensate for a loss of one camera?". For PLATO such analyses were performed as an example for three system parameters – the number of available cameras, the cadence and the TNID radiation level – to estimate their impact to the NSR.

PLATO is designed with 24 Normal Cameras. At EOL it is required that at least 22 cameras are still available (worst-case). The EOL scenario assumes that up to 2 cameras could be lost during the mission due to technical failures. This number results from a reliability analysis performed during the design phase.

The sensitivity analysis w.r.t. number of cameras shows what NSR to expect if the number of cameras is different and it depicts the fictive capabilities and limitations with even more cameras. The estimated NSR is related to stars with $m_V = 11$, all other parameters were kept constant. The required value is marked bold as reference. Figure 6 also allows to estimate the NSR for those stars which are covered just by 6, 12 or 18 cameras simultaneously (see green areas in Figure 1). The graphic also shows the values if none of the cameras would fail (still 24 cameras) as well as the numbers assuming 28 or 32 cameras.

| number of cameras | overall noise |
|---|---|
| [-] | [ppm] |
| 6 | 94,8 |
| 12 | 67,3 |
| 18 | 55,2 |
| 20 | 52,4 |
| **22** | **50,1** |
| 24 | 48,0 |
| 28 | 44,6 |
| 32 | 41,8 |

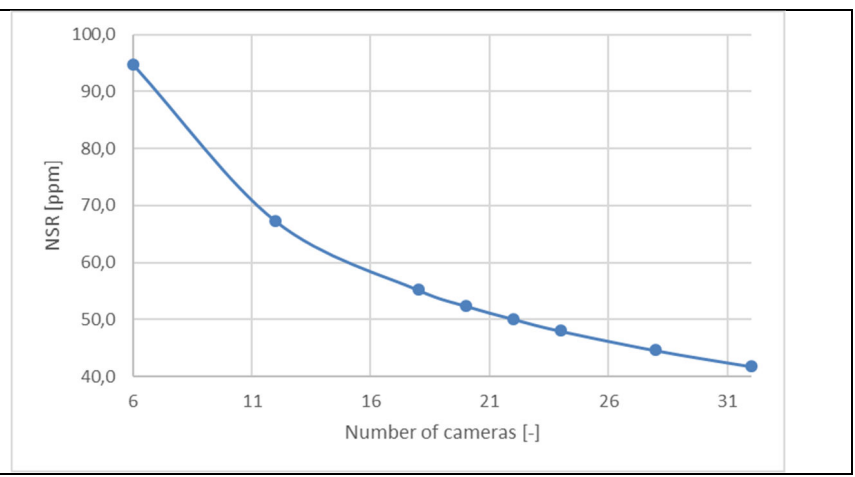

**Figure 6 Sensitivity analysis, NSR as a function of the number of available cameras, number of cameras greater than 24 is just for illustration.**

Cadence and exposure time will be the only major camera parameters that can be changed in space. Although increasing the exposure time, see Figure 7, would in principle improve the NSR, it would negatively affect the sampling frequency of the PLATO light curves and would also increase the number and degree of saturated stars within the FoV. This would also have a negative impact on the data product quality of bright stars too due to blooming effects. Moreover, the approach of increasing the exposure time assumes that spacecraft jitter noise will be lower than the requirement for longer time frames. This option might nevertheless be required in case of a possible loss of some additional cameras. If this would be the case, such a scenario should be investigated in detail to optimize the output.

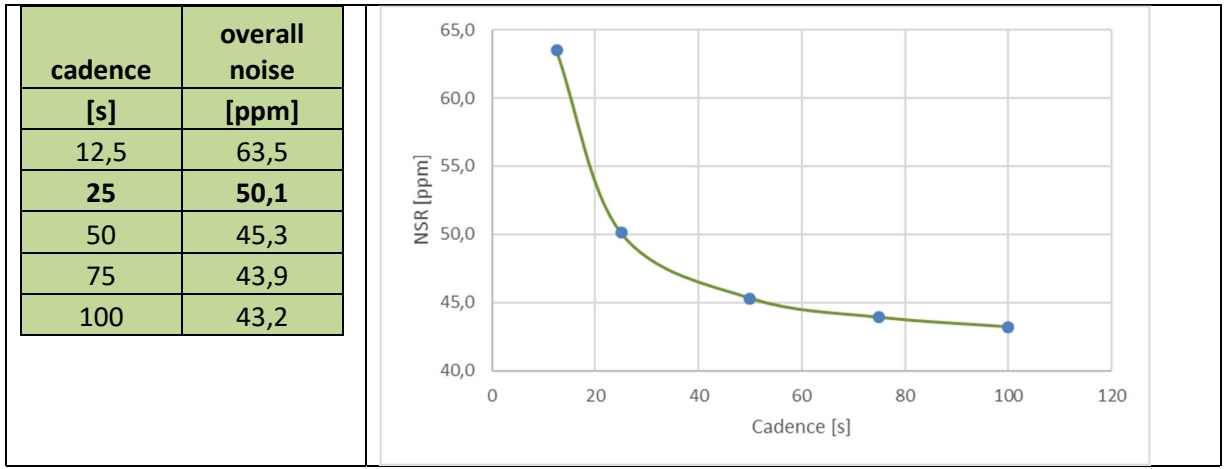

| cadence [s] | overall noise [ppm] |
|---|---|
| 12,5 | 63,5 |
| **25** | **50,1** |
| 50 | 45,3 |
| 75 | 43,9 |
| 100 | 43,2 |

**Figure 7 Sensitivity analysis, NSR as a function of the instrument's observing cadence**

Although there are precise radiation environment models, radiation in space cannot be predicted in a sufficient way mainly due to the random radiation characteristics of the sun. Since it is known that TNID will have a strong impact to the instrument's NSR, mainly due to CTI effects, a sensitivity analysis was performed (Figure 8). Effects are strongly non-linear and will have a different impact on various CCDs or even different regions in one CCD. This simulation shows the increase of the NSR with increasing TNID. In case of relevance, a scenario with strongly uneven impact on various CCDs has to be investigated in detail.

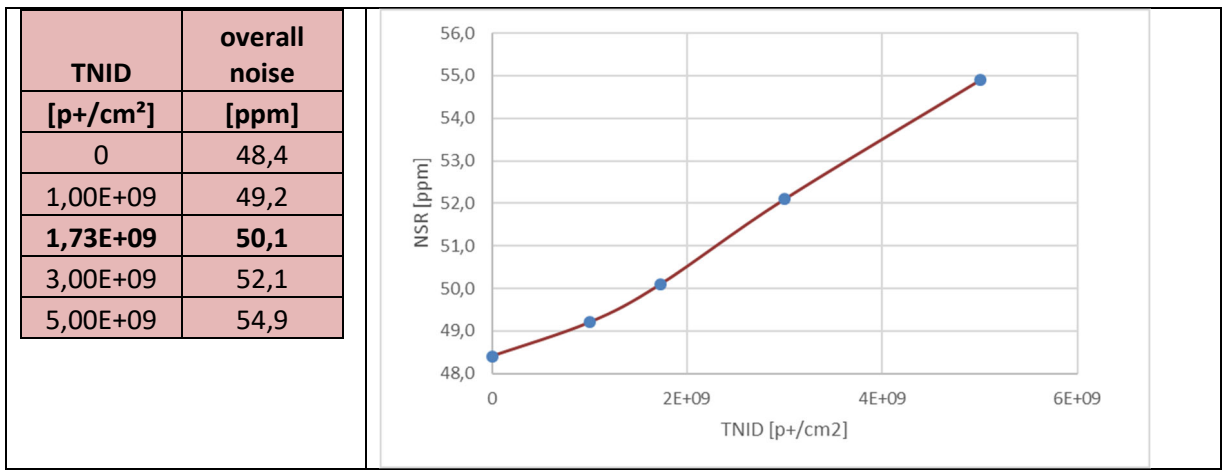

| TNID [p+/cm²] | overall noise [ppm] |
|---|---|
| 0 | 48,4 |
| 1,00E+09 | 49,2 |
| **1,73E+09** | **50,1** |
| 3,00E+09 | 52,1 |
| 5,00E+09 | 54,9 |

**Figure 8 Sensitivity analysis, NSR as a function of TNID radiation level**

## 6. Conclusion and outlook

The PINE simulator introduced here was used for a number of important investigations within the PLATO project in the past. The most important result of the analysis is that with the given system architecture, parameters and observation conditions, PLATO will be able to meet all requirements from beginning of life till end of life. The main noise contributor of a bright target star is, by design of the mission, the photon noise of the star itself. PINE is currently used in the consortium to define the different stellar samples (PLATO Input Catalogue, PIC, Montalto 2021) that will be observed by the mission.
Now, as PLATO is in the phase of manufacturing, assembly, integration and test, the underlying methodologies and tools can be used for different running tasks, e.g. cost-benefit analyses. PINE can be used with real parameter values before and even after launch, may they be 'worst case' or 'typical', allowing to get a more realistic picture of the expected performance.

In its medium mode PINE is able to handle spatial dependencies of camera parameters such as optical transmission and CTI. By doing so, NSR can be determined not only as an average value for the camera but for dedicated objects such as for pre-defined stars or for calibration targets. PLATO's star catalogue PIC was and is enriched by NSR values applying PINE in its medium mode. This approach will be published in a second paper.

## Acknowledgements

Gathering all the know-how, information and data needed for this paper and to keep them consistent was a tough job. It was a big and long-lasting system engineering task and is a masterpiece of collaborative work. Sincere thanks go to all colleagues and supporters from all the institutions and companies even if not mentioned by name.

This work presents results from the European Space Agency (ESA) space mission PLATO. The PLATO payload, the PLATO Ground Segment and PLATO data processing are joint developments of ESA and the PLATO Mission Consortium (PMC). Funding for the PMC is provided at national levels, in particular by countries participating in the PLATO Multilateral Agreement (Austria, Belgium, Czech Republic, Denmark, France, Germany, Italy, Netherlands, Portugal, Spain, Sweden, Switzerland, Norway, and United Kingdom) and institutions from Brazil. Members of the PLATO Consortium can be found at https://platomission.com/. The ESA PLATO mission website is https://www.cosmos.esa.int/plato. We thank the teams working for PLATO for all their work.

This work has been supported by the Belgian federal Science Policy Office (BELSPO) through the PRODEX grant "PLATO mission development" and by Spanish MCIN/AEI/10.13039/501100011033 grant PID2019-107061GB-C61.

J. Miguel Mas-Hesse is funded by Spanish MCIN/AEI/10.13039/501100011033 grant PID2019-107061GB-C61.

## Affiliations
1) Deutsches Zentrum für Luft- und Raumfahrt e.V. (DLR), Institut für Optische Sensorsysteme, Rutherfordstr. 2, 12489 Berlin, Germany, email: anko.boerner@dlr.de
2) Deutsches Zentrum für Luft- und Raumfahrt e.V. (DLR), Institut für Planetenforschung, Rutherfordstr. 2, 12489 Berlin, Germany
3) Centro de Astrobiologia (CAB), CSIC-INTA, 28692, Villanueva de la Cañada, Madrid, Spain
4) Istituto Nazionale di Astrofisica (INAF), Osservatorio Astrofisico di Catania, Via S. Sofia 78, 95123, Catania, Italy
5) European Space Agency (ESA)/ European Space Research and Technology Centre (ESTEC), Keplerlaan 1, 2201 AZ Noordwijk, The Netherlands
6) Istituto Nazionale di Astrofisica (INAF), Osservatorio Astronomico di Padova, vicolo dell'Osservatorio 5, 35122 Padova, Italy
7) OHB System AG, Manfred-Fuchs-Str. 1, D-82234 Wessling, Germany
8) Institute of Astronomy, KU Leuven, Celestijnenlaan 200D, B-3000 Leuven, Belgium
9) Laboratoire d'Etudes Spatiales et d'Instrumentations en Astrophysique (LESIA), Observatoire de Paris, Université PSL, CNRS, Sorbonne Université, Univ. Paris Diderot, Sorbonne Paris Cité, 5 Place Jules Janssen, 92195 Meudon, France
10) University College London (UCL), Mullard Space Science Laboratory (MSSL), Holmbury St. Mary, Dorking, Surrey, RH5 6NT, United Kingdom
11) Technische Universität Berlin, Zentrum für Astronomie und Astrophysik, Hardenbergstraße 36, 10623 Berlin, Germany
12) Freie Universität Berlin, Institut für Geologische Wissenschaften, Malteserstraße 74-100, 12249 Berlin, Germany

## Author's Contribution
Anko Börner wrote the main manuscript text. Carsten Paproth, Juan Cabrera, Martin Pertenais, Heike Rauer and J. Miguel Mas-Hesse provided essential contributions. All authors reviewed the manuscript.

## Competing Interests
To the best knowledge of the authors there are neither financial nor non-financial competing interests.